# Leaderboard Effects on Player Performance in a Citizen Science Game

Mads Kock Pedersen[1], Nanna Ravn Rasmussen[1], Jacob F. Sherson[1], and Rajiv Vaid Basaiawmoit[2]
[1]ScienceAtHome, Department of Physics and Astronomy, Aarhus University, Aarhus Denmark
[2]Faculty of Science and Technology, Aarhus University, Aarhus Denmark
madskock@phys.au.dk
nanna.phys@gmail.com
sherson@phys.au.dk
rvb@au.dk

**Abstract:** Points, badges, and leaderboards are some of the most often used game elements, and they are typically the first elements that people use from the gamification toolbox. Very few studies have tried to investigate the effects of the individual elements and even fewer have examined how these elements work together.

Quantum Moves is a citizen science game that investigates the ability of humans to solve complex physics challenges that are intractable for computers. During the launch of Quantum Moves in April 2016 the game's leaderboard function broke down resulting in a "no leaderboard" game experience for some players for a couple of days (though their scores were still displayed). The subsequent quick fix of an all-time Top 5 leaderboard, and the following long-term implementation of a personalized relative-position (infinite) leaderboard provided us with a unique opportunity to compare and investigate the effect of different leaderboard implementations on player performance in a points-driven citizen science game.

All three conditions were live sequentially during the game's initial influx of more than 150.000 players that stemmed from global press attention on Quantum Moves due the publication of a Nature paper about the use of Quantum Moves in solving a specific quantum physics problem. Thus, it has been possible to compare the three conditions and their influence on the performance (defined as a player's quality of game play related to a high-score) of over 4500 new players. These 4500 odd players in our three leaderboard-conditions have a similar demographic background based upon the time-window over which the implementations occurred and controlled against Player ID tags. Our results placed Condition 1 experience over condition 3 and in some cases even over condition 2 which goes against the general assumption that leaderboards enhance gameplay and its subsequent overuse as a an oft-relied upon element that designers slap onto a game to enhance said appeal. Our study thus questions the use of leaderboards as general performance enhancers in gamification contexts and brings some empirical rigor to an often under-reported but overused phenomenon.

# 1. Introduction

Games are an excellent field for studying behavioral theory and psychology, since games as an arena meets a very crucial criterion: We do it voluntarily. With a knowledge of why we play games, we can start using this, to design the world around us in such a way, that we start gaming our way through everyday challenges, a design process also known as gamification, which has been defined by Deterding (2011) as: "the use of game design elements in a non-game context". Understanding the psychology behind gaming is a complex matter, and trying to understand the complete nature of the human mind is today fully intractable. However, one can use at least a partial understanding to design concrete tools to use in the gamification industry.

Over the last couple of years, gamification has been used to improve motivation (Deterding, 2011), production (Mao et al. 2013), or learning (Pedersen et al. 2016) in research and educational contexts. The most common gamification approach has been to add a structure of points, badges and leaderboards around otherwise tedious tasks (Werbach and Hunter, 2012; Hamari et al, 2014). There have however been very few studies on the effect of the individual game-elements (Deterding, 2011; Hamari, Koivisto, and Sarsa, 2014; Seaborn and Fels, 2015), since they have been more concerned with the overall effect of introducing gamification, and primarily shown that there is a large dependence upon the particular combination and implementation of game-elements and the general context (Morschheuser, Hamari, and Kovisto, 2016).

Apart from increased corporate productivity, one of the most promising ideas risen from the process of gamification, is the concept of citizen science games (Lieberoth et al, 2014; Kullenberg and Kasperowski, 2016), especially after games like FoldIt (Cooper et al, 2010), EteRNA (Lee et al, 2014), and EyeWire (Kim et al, 2014) enabled concrete scientific results obtained by players. The idea of making a game to get people to solve science problems introduces a completely new way of producing knowledge via Citizen Science. With scientists venturing into the field of making games - usually with a smaller budget and less experience than commercial companies - it becomes very relevant to question how the various implementations work in different contexts and how the game-elements interact with each other, since the various solutions can have a markedly different development costs and run-time resource requirements.

Leaderboards are one of the most commonly used game-elements - often in connection with points - and leaderboards can be implemented in a multitude of different ways depending on the context, and the specific motif behind including them; increase motivation (Halan et al, 2010), increasing competition (Landers and Landers, 2015), or plainly setting a goal to the individual player (Landers, Bauer, and Callan, 2015). In general leaderboards can roughly be divided into two different categories: Arcade style top N leaderboards that only displays the top players regardless of the current player's position on it, and relative leaderboards that display a player's current position on the leaderboard and the other players just above or below on the leaderboard. Beyond this crude classification there are a lot variants that can be mixed and combined e.g. scrollable leaderboards, local leaderboards, time limited leaderboards, or multi-level leaderboards. However, the actual effects of leaderboards on players is still poorly understood and there exists a paucity in the literature around this topic. It is necessary to investigate the basic properties of leaderboards in general, before we can even begin to map the intricate details of the more complex leaderboards (Dicheva et al, 2015). Previous studies that have attempted to isolate leaderboard effects have shown that leaderboards rewarding quantity leads to a higher production (Farzan et al, 2008; Mekler et al, 2015; Wu, Kankanhalli, and Huang, 2016), while the effect on quality is mixed (Jung, Schneider, and Valacich, 2010; Dominquez et al, 2013; Landers, Bauer, and Callan, 2015). Other studies have looked on the effect of the leaderboards over time and have found that leaderboards are most efficient on a short time scale where obtaining a top position is possible (Massung et al, 2013; Ipeirotis and Gabrilovich, 2014)

Quantum Moves is a citizen science game designed by quantum physicists at Aarhus University wanting to solve a complex stumbling block in the process of designing a quantum computer (Lieberoth et al, 2014). In the game, the player moves an atom around in a potential well, facing different tasks with different levels of difficulty. The data created by the players are then improved using a local search computer algorithm. This combined human-computer search yielded solutions far superior to those obtainable by the computer algorithms alone (Sørensen et al, 2016).

As quantum moves was launched on large scale to the public, a great opportunity to study some of the theories behind game design in a practical setting came along, when the leaderboard was changed several times. In this study, we therefore present a test of different leaderboard types (Top 5, relative, and none) and their effect on the players of Quantum Moves. This constitutes a rough, though directly applicable test of the leaderboard as a motivator for continuous playing.

## 2. Data collection

On April 12th 2016 Quantum Moves was launched on Apple AppStore and Google Play, and at the same time made available for download to Windows, Mac, and Linux through quantummoves.org. This was timed to coincide with the publication of a paper in Nature by Sørensen et al. (2016) reporting on the human-computer results obtained from the previous version of the game. The findings of the Nature paper were reported in more than 100 news media outlets across the world and as a result of this massive exposure over a two week period Quantum Moves had 150.000 new players playing it 5 million times. All playthroughs of the games were saved in their entirety in order to allow for later data analysis.

Quantum Moves - see Figure 1 - is a single player game in which the player has to complete a series of levels with increasing difficulty. In each level the user can be awarded up to 3 stars depending upon the quality and speed of the solution. This provides an incentive for the user to replay the level and improve upon his/her own score. Another feature in the game that is supposed to incentivize replaying levels was leaderboards. At the time of the launch there was a relative leaderboard that displayed the user's highscore in a specific level together with the two users just above and the two users just below in the rankings, but the technical implementation was turned out not to be able to scale to the large amount of traffic. As a consequence the leaderboard were slow to load on the players' screens, and by April 14th in the morning it did not load at all, which effectively left Quantum Moves without a leaderboard. A new implementation of the relative leaderboard that would scale with the traffic proved to be difficult to implement, thus a temporary fix introduced a Top 5 leaderboard per level from April 16th 12:47 am until April 26th 15:05 pm where the relative leaderboard was reinstated for good. An example of the relative leaderboard can be seen in Figure 1, the Top 5 leaderboard had a similar look, while the faulty leaderboard showed all the same things except for the ranking.

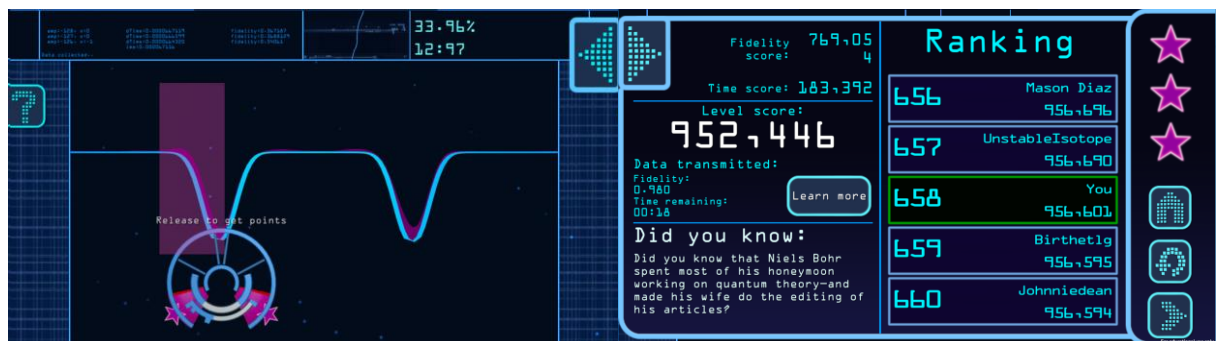

**Figure 1.** Left: Screen shot during playthrough of Quantum Moves. The blue line represents the potential energy landscape, while the purple fluid on the top of the potential is the representation of the atom. Right: Screenshot of the relative leaderboard from Quantum Moves.

These changes in the leaderboard implementations and the large influx of new players give a unique opportunity to test leaderboard effects in a points driven citizen science game, since the other conditions were the same. The results section will present data obtained in three separate 6 hour periods, each period is asociated with a specific leaderboard implementation and begins at the first midnight after the associated leaderboard were in effect. We will only compare data from new players and their playthroughs within the time windows and we label these "conditions" as follows:

1. *No leaderboard:* The leaderboard was broken, and the only thing people saw was a loading-error and their own score. Data was collected from April 15th 00:00:01 to April 15th 06:00:01
2. *Top 5 leaderboard*: The players could see the top 5 players. The players could see their own score, but not their rank in the game. Data was collected from April 17th 00:00:01 to April 17th 06:00:01

3. *Relative leaderboard*: Showed the rank of the individual player and the two nearest players with lower and higher ranks respectively. Data was collected from April 27th 00:00:01 to April 27th 06:00:01

## 3. Results

During the three data collection periods, we had a total of 4553 new players (No leaderboard: 1875 players, Top 5: 2124 players, and relative leaderboard: 554 players). There were fewer people in the relative leaderboard condition since this data was collected on the tail-end of the media-hype surrounding the publication of the Nature article.

The first thing we looked into was if the leaderboards led to a higher degree of replays of the game (see Table 1). We defined a replay as a player having a break in gameplay of *at least* 30 minutes starting within the condition window before resuming playing the game.

**Table 1:** Replays from the different leaderboards condition. To register the replays we looked at all the recorded data in Quantum Moves until June 4th 2016.

|  | No Leaderboard | Top 5 leaderboard | Relative leaderboard |
|---|---|---|---|
| **Players replaying the game** | 8.6% | 7.3% | 5.6% |

In order to distinguish whether the conditions were significantly different we performed Fisher's exact test with a p<0.05 significance level and the Bonferroni correction i.e. significance level for each test is *p*<0.017. The test did not reveal any significant differences.

We have looked into how large a fraction of players within the conditions that played any given level (see Figure 2). We do see that the Top 5 leaderboard systematically retains the highest fraction of players, but we have not been able to find a statistic significance of this.

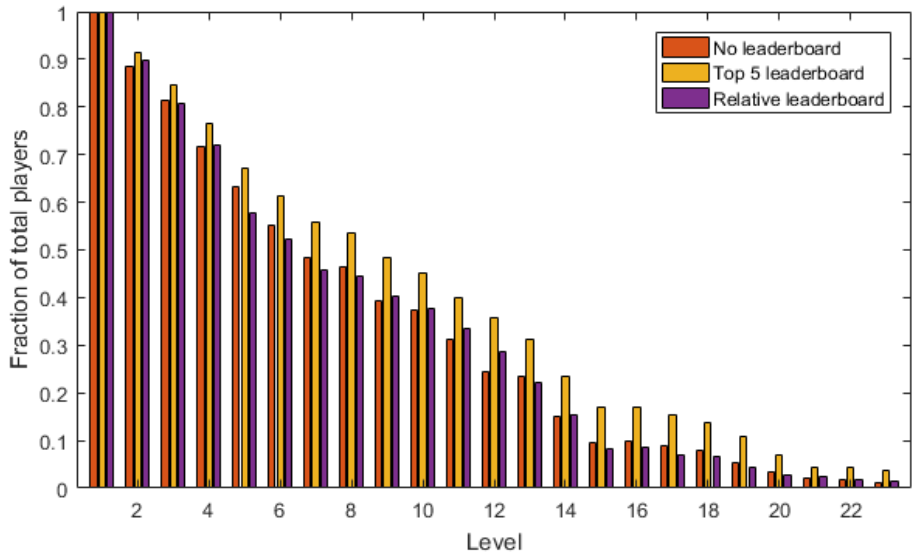

**Figure 2:** Fraction of players within each leaderboard condition that played a given level.

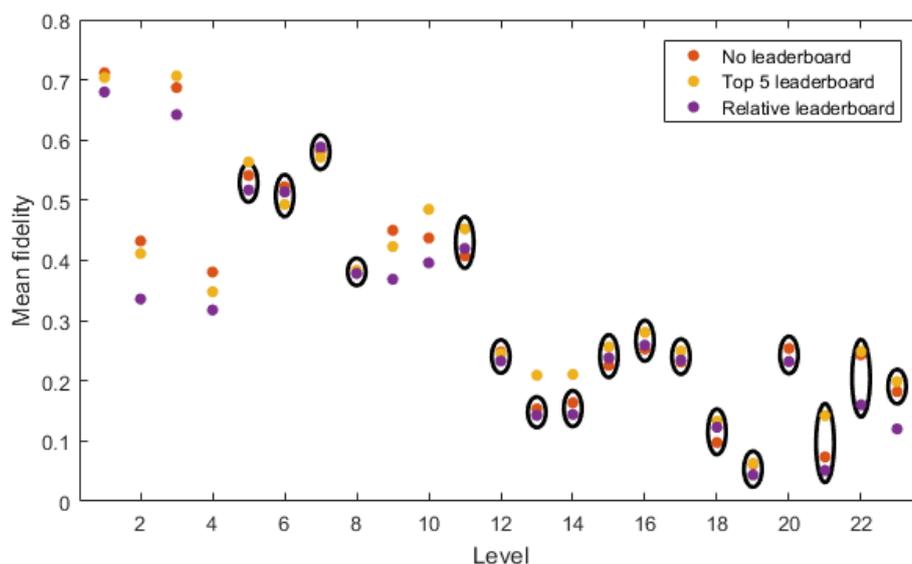

**Figure 3:** Mean fidelity within each of the leaderboard conditions for a given level. The conditions that were not significantly different from each other have been grouped together by a black ring.

In order to tell if any of the conditions produced better solutions for the scientific investigations we look at the fidelity of a solution, i.e. how much does the atom at the end of the playthrough resemble the desired state (see Sørensen et al. 2016 for a detailed description of how the fidelity of a solutions is calculated). We then made a comparison of all the solutions generated for a specific level (see Figure 3) between the different conditions by performing a Kruskal-Wallis test. For levels 1, 2, 3, 4, 5, 9, 10, 13, 14, and 23 we found a significant difference between the conditions, which we further resolved with a post-hoc Mann-Whitney U test. The conditions that could not be significantly distinguished from each other have been grouped together with a black ring in Figure 3.

## 4. Discussion

The degree of generalizability of case studies like these depends critically on a thorough discussion of the particular settings within which the results are obtained. Points, badges, and leaderboards can be used in many different contexts/framings, which will make a particular observation more or less likely. In the following, we therefore carefully review the intention and the details of the implementation points, badges, and leaderboards. As mentioned above they are three of the most often seen elements in gamification, which have also led to the assumption that these elements are essential elements within any game design. Points, an element for keeping score and a "relative marker" for your skill level, are one of the most used elements for providing real-time feedback of one's accomplishment in a game. In Quantum Moves, points are scored based on the quality and speed of the solution generated. The higher the points – the closer you are to having moved the atom "the right way" – which we match with real-world conditions using the "fidelity" parameter as described in the Results section. Besides factoring into points, the fidelity also triggers certain thresholds that then allow you to gain stars (i.e. badges). There are three stars that you can attain – which serves a dual purpose – first, they give players a real-time feedback of their solution, which incentivizes them to re-play the level to try to earn all three stars before they can progress to the next level. This helps overall quality and allows the researchers to get more high-quality solutions that are fairly reproducible, which helps the computer algorithms fine-tune the player data through a process of "optimization". Secondly, the badges also act as an intrinsic motivation tool that helps in player retention and contributes to a sense of accomplishment for the player. Last, but not the least, points are also the main way through which your position on the leaderboard is decided. Badges – can therefore be called "achievement markers" but unlike points, which need context (other players points or max/min levels to give you an idea of where you stand), badges are able to give you a sense of achievement, skill level, accomplishment without much context or feedback from the system or other players. While badges are often used as self-measures and give players a sense of accomplishment as compared to themselves only, badges can also be used in social context. For example, in a popular game like "Candy Crush", badges are used both as a skill marker but also as a social status marker – i.e.

do you have more stars than your friends? In Quantum Moves, the stars are only visible to local players – thus they only serve the process of self-reflection devoid of any visible social/peer pressures.

The social element in quantum moves is most visible through the leaderboard element, the third element in the popular points, badges, and leaderboards lexicon. Leaderboards rank a player on a board and are considered a real-time, dynamic ranking list that also acts as a social status marker (Zichermann and Linder, 2013). Wang and Sun (2011) also found that leaderboards are used for self-assessment in long-term progress in a game. Nevertheless, Leaderboards are not directly a social element – i.e. they do not facilitate social interaction *per se* but can act as triggers for social interaction and engagement as social status markers. Leaderboards also set a goal in terms of achievement attainment that players can strive towards as it indicates that if someone else has gotten a higher score than you, then essentially that position on the leaderboard is also accessible to you if you are willing to strive for it. These reasons combined are probably one of the reasons that leaderboards are so prevalent in many games. In the design of Quantum Moves, the decision to add leaderboards to the game was also based upon this preceding information.

Gamification experts have often suggested that overuse or wrong use of Leaderboards can drive players contrary to expectations. Zichermann and Lindner (2013) have found that while leaderboards are motivating to some, they can be demotivating to others. Due to this polarization in terms of motivation, Zichermann and Lindner (2013) and many other gamification experts have suggested that leaderboards be individualized or offer relative leaderboards that can allow those at the bottom to compare themselves with those closer to them and thus removing the frustration that often arises with players limited movement on leaderboards. Thus, over the last few years there has been an increase in the implementation of relative leaderboards across a broad gamut of games (both commercial and non-commercial citizen science games). Whether this is enough to convince players who are put off by competition or do not find leaderboards interesting is still an open question.

Additionally, relative leaderboards are technically more challenging to setup and require considerably more resources to set up. Nevertheless, the perceived gain from leaderboards is why ScienceAtHome set up leaderboards and then relative leaderboards in Quantum Moves. While not specifically setting up an experiment to test out the impact of leaderboards, the breakdown in our implementation while the game was live, offered a unique insight into a phenomenon that is widely accepted but less studied.

Our results, in general, questions the use of leaderboards or rather the over-reliance of leaderboards and especially the relative leaderboard – which was introduced to counter the perceived negative effect of the Top 5 leaderboard type of setup. Figure 2 shows that the Top 5 leaderboard consistently retains the highest fraction of players, but what is even more interesting is that the fraction of players retained in the no leaderboard and relative leaderboard conditions are equal. It should be noted that since the both Top 5 leaderboard and the relative leaderboard retained all the previous high-scores – including those outside the data collection windows – it was relatively hard to obtain a top position. However, incremental improvements in the score led to huge improvements in the absolute player ranking. No differences does not mean no result – but rather that due to the many factors that influence the final results it indicates that this area has to be investigated more thoroughly and our study offers the first high-power empirical insight into the impact or the lack of impact of leaderboards.

Figure 3 reaffirms our observations, when taking into account player high score and quality of data produced, and the data is especially more revealing at the easier levels of the game where the differences seen in the 3 conditions are statistically relevant. An average scan of the data here too indicates that the relative leaderboard is consistently trailing across all levels – whereas no leaderboards and Top 5 leaderboard condition equally share levels where they hold the top place in terms of fidelity. This too questions the reliance of games on leaderboards.

In terms of citizen science games, this insight becomes even more relevant. As opposed to commercial games that are driven by commercial incentives, that is another bias (commercial drivers intentionally driving behavior to aid in commercial exploitation) we are able to eliminate by having a non—commercial, yet popular game (in terms of number of players having played the game). Secondly, where commercial games have more resources at their disposal, citizen science games are often limited in terms of access to resources. This requires a more judicious use of available resources and our data suggest that unless there is a significant gain that can be had and **documented** by applying a relative leaderboard – relative leaderboards do not justify resource use in a purely cost-benefit perspective.

## 5. Conclusion

The data collected in three different leaderboard conditions (no leaderboard, Top 5 leaderboard, and relative leaderboard) from the citizen science game Quantum Moves, have not revealed any strong preferences for a particular leaderboard implementation in terms of retention and performance, although the study was performed with an unusually high power for leaderboard studies. We have, however, not looked into player attitudes towards the different leaderboard implementations, neither whether there are specific subgroups of players for which the leaderboard had a substantial effect. The lack of significant differences could also be due to the nature of the game play in Quantum Moves, and that other types of game play would lend themselves to be driven more by leaderboards. Thus, further studies with designed and controlled experiments would be in order to cast light on these issues.

In conclusion, leaderboards do have a place in both commercial and non-commercial games. They can provide players with feedback upon their progression, set a goal and incentivize competition. Thus, our data should not be used/considered as an argument for completely abandoning them, but rather serve as a good reminder that leaderboards are not a panacea that automatically makes a game fun or more engaging. Rather, one would need to consider carefully how and when leaderboards fit into the games, which leaderboard option to consider and for what purpose. Overall, our results should lead to more sensible game-design decision-making as well as aid low-resource citizen science initiatives to better prioritize the limited resources at their disposal when developing and designing games as tools for their project.

**Acknowledgements**
Financial support from the Lundbeck Foundation, the Carlsberg Foundation, and the John F. Templeton Foundation is gratefully acknowledged.

**References**
Cooper, S., Khatib, F., Treuille, A., Barbero, J., Lee, J., Beenen, M., Leaver-Fay, A., Baker, D., Popović, Z., and Foldit players (2010). Predicting protein structures with a multiplayer online game, *Nature* Vol. 466, pp. 756–760

Deterding, S. (2011). *Situated motivational affordances of game elements: a conceptual model*. In CHI 2011 Workshop Gamification: Using Game Design Elements in Non-Game Contexts.

Dicheva, D., Dichec, C., Agre, G., and Angelova, G. (2015), Gamification in Education: A Systematic Mapping Study. *Journal of Educational Technology & Society* Vol. 18, pp. 75

Dominguez, A., Senz-de-Navarrete, J., de-Marcis, L., Ferández-Sanz, L., Pages, C., and Martínez-Herraiz, J.-J. (2013). Gamifying learning experinces: pratical implications and outcomes. *Computers & Education*, Vol. 63, pp 380

Farzan, R., DiMicco, J.M., Millen, D.R.M.B, Dugan, C., Geyer, W., Brownholtz, E.A. (2008). *Results from deploying a participation incentive mechanism within the enterprise*. In Proceedings of CHI 2008, pp. 563-572

Halan, S., Rossen, B., Cendan. J., Lok, B. (2010). *High Score! - Motivation Strategies for User Participation in Virtual Human Development*, In: Allbeck J., Badler N., Bickmore T., Pelachaud C., Safonova A. (eds) Intelligent Virtual Agents. IVA 2010. Lecture Notes in Computer Science, Vol 6356. Springer, Berlin, Heidelberg

Hamari, J., Koivisto, J., and Sarsa, H. (2014). *Does Gamification Work? - A Literature Review of Empirical Studies on Gamification*. In Procedings of the 47th Havaii International Conference on System Sciences. pp 3025

Ipeirotis, P.G. and Gabrilovich, E. (2014). *Quizz: Targeted Crowdsourcing with a Billion (Potential) Users*. Proceedings of the 23rd international conference on World wide web - WWW '14, Seoul, ACM, pp. 143

Jung, J.H., Schneider, C., and Valacich, J (2010). Enhancing the motivational affordance of informations systems; the effects of real-time performance feedback and goal setting in group collaboration environments. *Management Science*, Vol. 56, pp 724

Kim, J. S., Greene, M.J., Zlateski, A., Lee, K., Richardson, M., Turaga, S.C., Purcaro, M., Balkam, M., Robinson, A., Behabadi, B.F., Campos, M., Denk, W., Seung H.S., and the EyeWirers (2014). Space-time wiring specificity supports direction selectivity in the retina. *Nature* Vol. 509, pp. 331–336

Kullenberg, C., and Kasperowski, D. (2016). What Is Citizen Science?–A Scientometric Meta-Analysis. *PLOS ONE*, Vol 11, pp. e0147152

Landers, R.N., Bauer, K.N., and Callan, R.C. (2015). Gamification of task performance with leaderboards: A goal setting experiment. *Computers in Human Behavior*, Vol. 71, pp 508

Landers, R.N. and Landers, A.K. (2015). An empirical test of the theory of gamified learning: the effect on time-on-task and academic performance. *Simulation & Gaming*, Vol 45, pp 769

Lee, J., Kladwang, W., Lee, M., Cantu, D., Azizyan, M., Kim, H., Limpaecher, A., Gaikwad, S., Yoon, S., Treuille, A., Das, R., and EteRNA Participants (2014). RNA design rules from a massive open laboratory. *PNAS* Vol. 111, pp. 2122–2127

Lieberoth, A., Pedersen, M. K., Marin, A., Planke, T., and Sherson, J. F. (2014). Getting Humans to Do Quantum Optimization - User Acquisition, Engagement and Early Results from the Citizen Cyberscience project Quantum Moves. *Human Computation*, Vol. 1, No. 2, pp 221

Mao, A., Kamar, E., Yuling, C., Horvitz, E., Schwamb, M.E., Lintott, C.J., Smith, A.M. (2013). *Volunteering Versus Work for Pay: Incentives and Tradeoffs in Crowdsourcing.* In Proceedings of the First AAAI Conference on Human Computation and Crowdsourcing, Palm Springs, pp. 94

Massung, E., Coyle, D., Cater, K.F., Jay, M., and Preist, C. (2013). *Using crowdsourcing to support pro-environmental community activism*. In Proceedings of the SIGCHI Conference on Human Factors in Computing Systems - CHI '13, Paris, pp. 371–380.

Mekler, E.D., Brühlmann, F., Tuch, A.N., and Opwis, K. (2015). Towards understanding the effects of indivdual gamification elements on intrinsic motivation and performance. *Computers in Human Behavior*, Vol. 71, pp 525

Morschheuser, B., Hamari, J., and Kovisto, J. (2016). *Gamification in Crowdsourcing: A Review*. In Proceedings of the 49th Hawaii International Conference on System Sciences. pp. 4375

Pedersen, M. K., Skyum, B., Heck, R., Müller, R., Bason, M. Lieberoth, A., and Sherson, J. F. (2016) Virtual Learning Enviroment for Interactive Engagement with Advanced Quantum Mechanics. *Physical Reveiw Physics Educuation Research*, Vol. 12, pp 013102

Seaborn, K. and Fels, D.I. (2015). Gamification in theory and action: a survey. *International Journal of Personality and Social Psychology*, Vol. 74, pp 14

Sørensen, J. J. W. H., Pedersen, M. K., Munch, M., Haikka, P., Jensen, J. H., Planke, T., Andreasen, M. G., Gajdacz, M., Mølmer, K., Lieberoth, A. and Sherson, J. F. (2016) Exploring the quantum speed limit with computer games, *Nature*, Vol. 532, No. 7598, pp 210

Wang, H., & Sun, C. T. (2011). *Game reward Systems: gaming experiences and social meanings.* In Proceedings of the DiGRA 2011 Conference: Think design play

Werbach, K., and Hunter, D. (2012). *For the Win: How Game Thinking Can Revolutionize Your Business*. Wharton Digital Press

Wu, Y., Kankanhalli, A., Huang. K.-W. (2016). *Gamification in Fitness Apps: How Do Leaderboards Influence Exercise?* In Proceedings of 36th International Conference of Information Systems, Fort Worth, USA.

Zichermann, G. and Linder, J. (2013). *The Gamification Revolution: How Leaders Leverage Game Mechanics to Crush the Competition*. McGraw-Hill Education, Columbus, USA